\documentclass[twocolumn,showpacs,preprintnumbers,amsmath,amssymb]{revtex4}
\usepackage{epsfig}
\usepackage{graphicx}
\usepackage{dcolumn}
\usepackage{bm}
\begin{document}

\title{Supersymmetry theory of microphase separation in
homopolymer--oligomer mixtures}

\author{Alexander Olemskoi}
\email{olemskoi@ssu.sumy.ua}
\author{Ivan Krakovsky}
\email{ivank@kmf.troja.mff.cuni.cz}
\author{Alexey Savelyev}
\email{alexsav@kmf.troja.mff.cuni.cz} \affiliation{Physical and
Technical Department,
Sumy State University, Rimskii-Korsakov St. 2, 40007 Sumy, Ukraine\\
Department of Macromolecular Physics, Charles University\\
V Hole\v{s}ovi\v{c}k\'{a}ch 2, 180~00 Prague 8, Czech Republic}

\date{ \today }

\begin{abstract}
Mesoscopic structure of the periodically alternating layers of
stretched homopolymer chains surrounded by perpendicularly
oriented oligomeric tails is studied for the systems with both
strong (ionic) and weak (hydrogen) interactions. We focus on the
consideration of the distribution of oligomers along the
homopolymer chains that is described by the effective equation of
motion with the segment number playing the role of imaginary time.
Supersymmetry technique is developed to consider associative
hydrogen bonding, self--action effects, inhomogeneity and
temperature fluctuations in the oligomer distribution. Making use
of the self--consistent approach allows to explain experimentally
observed temperature dependence of the structure period and the
order--disorder transition temperature and period as functions of
the oligomeric fraction for systems with different strength of
bonding. A whole set of parameters of the model used is found for
strong, intermediate and weak coupled systems being
P4VP--(DBSA)$_x$, P4VP-(Zn(DBS)$_2$)$_x$ and P4VP--(PDP)$_x$,
respectively. A passage from the formers to the latters shows to
cause crucial decrease of the magnitude of both parameters of
hydrogen bonding and self--action, as well as the order--disorder
transition temperature.
\end{abstract}

\maketitle

\section{Introduction}

Surfactant--induced mesomorphic structures based on the
association between flexible homopolymers and head--functionalized
oligomers represent a new class of supramolecular materials. They
exhibit a rich phase behavior to be an object of investigations
that have attracted, during past decade, considerable attention of
both experimentalists \cite{4}~--~\cite{2} and theoreticians
\cite{26}, \cite{27}. Microphase separation is the principal
property of such systems which results in the formation of ordered
mesoscopic structures due to the association between the head
group of the oligomer and corresponding groups of the homopolymer,
on the one hand, and unfavorable polar--nonpolar interactions
between the non--polar tail of the surfactant molecules and the
rest of the system, on the other one.

\begin{figure}[!h]
\epsfig{file=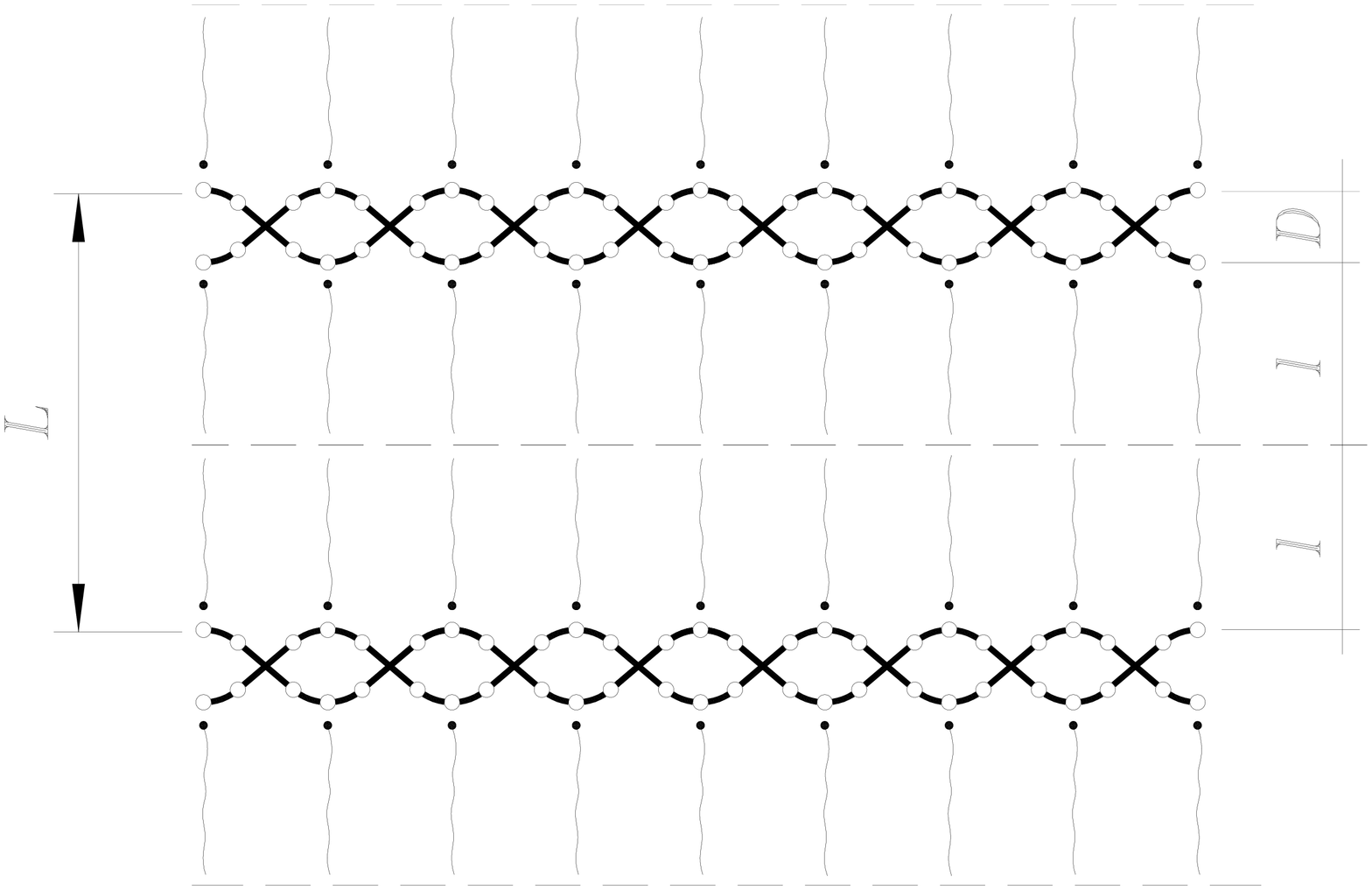,height=75mm,width=55mm,bbllx=17mm,bblly=37mm,bburx=183mm,bbury=291mm,angle=90}
\caption{Schematic picture of the homopolymer--oligomer microphase
separated structure for $x=1/3$.} \label{structure}
\end{figure}

The homopolymer--oligomer systems involve two main classes that
are relevant to strong ionic bonds and weak hydrogen ones. Unlike
to conventional copolymers where repulsive blocks are bonded
together by covalent bonds, there are various temporary physical
interactions which play a crucial role in the formation of ordered
mesophases in such systems. In the ionic bonding systems the
degree of association is relatively high, so that the polymer
chain resembles a comb copolymer with regularly alternating
oligomer side chains. At the same time, for the systems with
temperature--dependent hydrogen bonds the incompatibility must not
be so strong to induce separation on a macroscopic level. Here,
the microphase separation results in the periodic alternation of
the layers of stretched homopolymer chains surrounded by
perpendicularly oriented oligomer tails (see Figure
\ref{structure}).
Similarly to the conventional copolymer systems, a rich variety of
morphologies (lamellar, cylindrical, spherical etc.) is shown to
be possible \cite{4}. However, for the sake of simplicity we will
restrict ourselves with considering lamellar morphology only.

An example of the ionic bonding systems is represented by the
mixture P4VP--(DBSA)$_x$ of the homopolymer being atactic
poly(4-vinyl pyridine) (P4VP) and the surfactant as dodecyl
benzene sulfonic acid (DBSA). Here, owing to the very strong
interaction, the temperature domain of microphase separation is
not bounded from above by association effects \cite{4,DbsaZn}. The
peculiarity of the systems of this type, being
polyelectrolyte--surfactant complexes, is that the long space
structure period is an increasing function of the number
oligomer/monomer ratio $x$ (the number of DBSA--groups per one
pyridine ring). 
More complicated behavior is inherent in the hydrogen bonded
systems which were considered to study the opposite weak-bonding
limit \cite{Hb}~--~\cite{2}. Here, the weak interaction causes an
order--disorder transition to homogeneous high--temperature state.
An example of these systems is given by the mixture
P4VP--(PDP)$_x$ of the same homopolymer P4VP with 3-pentadecyl
phenol (PDP) being the oligomer. In this case, unlike to the ionic
bonding systems, the long space period decreases with
$x$--increase. An intermediate behavior exhibit the system
P4VP-(Zn(DBS)$_2$)$_x$ with the oligomer being zinc dodecyl
benzene sulfonate Zn(DBS)$_2$ that forms transition metal
coordination complexes with the monomers of P4VP \cite{DbsaZn}.
Ionic bond weakening due to the absence of covalently bound
charges along the homopolymer chain leads here to a non--monotonic
form of the $x$--dependence of the long space period.

Principally important for our consideration is decreasing form of
the temperature dependence of the long space period for all above
systems \cite{DbsaZn} -- \cite{2}. However, such character of the
dependence appears in hydrogen bonded systems only within a finite
temperature interval bounded by the glass transition temperature
$T_g$ from below and order--disorder transition temperature $T_c$
from above \cite{Hb,8}. Here, an increase of the oligomer/monomer
ratio $x$ leads to a non--monotonic behavior of the temperature
$T_c$ with a maximum near the point $x=0.85$, deviation from which
narrows the temperature domain $T_g\div T_c$. This domain is the
region of our interest where a purely microphase separated
structure is possible. Below the glass transition temperature
$T_g$ the crystallization of the oligomer chains occurs that
causes a reduction of the overall volume of the system and a
sudden decrease of the long space period \cite{8}.

Microphase separation phenomenon had been extensively studied  in
the past two decades for a variety of polymer systems including
random heteropolymers \cite{21} -- \cite{19}. Theoretical studies
of the homopolymer-oligomer mixtures, being the systems of
associating polymers, were proposed by Tanaka {\it et al.}
\cite{26} and Dormidontova {\it et al.} \cite{27} within the
random phase approximation introduced by Leibler \cite{21}. Here,
the total free energy
\begin{equation}
F_{tot}=F_{ho}+F_{hb} \label{000}
\end{equation}
is written as a sum of two terms, $F_{ho}$ related to the
non--associated homopolymer--oligomer mixture and $F_{hb}$
attributed to the hydrogen bonding. Then, making use of
minimization principle with respect to the dependence of the free
energy $F_{tot}$ on the average fraction of hydrogen bonds $X$
present in the system, permits to find the temperature dependence
$X(T)$ and to study possible forms of phase diagrams for both
macrophase and microphase separations. It turned out that this
approach gives the real dependence of the long space period $L$ of
the ordered structure on the oligomer/monomer ratio in the system,
however, as the fraction of hydrogen bonds monotonically decreases
with temperature, the increasing temperature dependence of $L(T)$
obtained is in contradiction to the experimental data \cite{8}.
This inconsistency is caused obviously by the roughness of the
random phase approximation used for description of the hydrogen
bonding.

To avoid this limitation, our approach is based on the above
mentioned analogy between associating homopolymer--oligomer
mixtures and random comb copolymers taking into account the
varying number of oligomers attached to the main chain
stochastically. Such a system can be analysed in terms of the
random walk statistics to apply the field theoretical scheme
\cite{10} for the development of the microscopic theory. The
corner stone of our approach lies in the assumption that the
alternation of the homopolymer associative groups with and without
oligomers attached is like the alternation of the segments of
different types along the chains of a random heteropolymer to be
represented as a stochastic variation of the Ising spin, for which
the role of imaginary time is played by the number of chain
segment $n$ \cite{11} -- \cite{13}.

Along this line, the problem under consideration is divided into
two parts, the first of which is reduced to the determination of
the relation between the long space period $L$ and the average
fraction of hydrogen bonds $X$, whereas the second one is focused
on the determination of the frequency $\omega=2\pi X$ in the
distribution of the oligomer heads along the homopolymer chain.
The first part of the problem was studied on base of the simplest
model \cite{2} that is reduced to the treatment of the dependence
$F_{ho}(L)$ given by the first term of the free energy
(\ref{000}). Corresponding consideration developed within the
framework of the strong segregation limit derives to generic
relation (\ref{1h}) for the dependence $L(\omega)$ (see Appendix
A). In this paper, we focus on the second problem to be related to
the definition of an optimal frequency $\omega$ that minimizes the
second term of the free energy (\ref{000}) within the framework of
the weak segregation limit.

The formal basis of our treatment is the field theoretical scheme
of stochastic systems with using the supersymmetry field
\cite{10}. Conformably to the polymers, this theory was proposed
in \cite{Vilgis} and developed for the random copolymers in Refs.
\cite{11}~--~\cite{13}. Our approach is based on the
Martin--Siggia--Rose method of the generating functional
\cite{Rose}. Power and generality of the supersymmetry field
scheme was demonstrated for the Sherrington--Kirkpatrick model for
which it is identical to the replica approach \cite{Kurchan}. The
formal base of the supersymmetry is a nilpotent quantity which
represents a square root of $0$. In this sense, the superfield is
similar to the complex field, in which the imaginary unit, being
square root of $-1$, is used instead of the anticommuting
nilpotent quantity being the Grassmann variable. By definition,
the supersymmetry field combines commuting the boson and
anticommuting fermion components into the unified mathematical
construction representing a vector in the supersymmetry space.
Choice of the optimal basis of the supersymmetry correlator yields
in optimal way the advanced/retarded Green functions and the
structure factor to obtain microscopic expression for the
frequency $\omega$.

The paper is organized in the following manner. Section II
contains initial relations of the field scheme used to write the
system Lagrangian. It involves the effective potential energy
whose quadratic term describes hydrogen bonding between the
oligomers and the associative groups of the homopolymer chains,
whereas cubic and biquadratic terms relate to the self--action
effects. The principal peculiarity of our approach lies in
accounting of the inhomogeneity in the distribution of oligomers
along the homopolymer chains. This accounting is caused by the
introduction of the effective kinetic energy whose density is
proportional to the square of the derivative of oligomers
distribution over segment numbers $n$. Due to the temperature
dependence of the hydrogen coupling, related effective mass is a
fluctuating parameter whose averaging, along the
Hubbard--Stratonovich procedure, arrives at the biquadratic term
with respect to the time derivative. According to the calculations
given in Section IV, just this term, being considered within the
mean--field approach, causes decaying character of the temperature
dependence $L(T)$ of the structure period. Complication of the
problem arising from the determination of the proper frequency
$\omega$ is caused by an essential non--linearity and coupling the
advanced/retarded Green functions and the structure factor. Hence,
it is methodically convenient to use the supersymmetry technique
that enables to obtain in the simplest way explicit expressions
for above functions in the long--range limit (see Section III).
Divergency condition of the Green function permits to find the
proper frequency $\omega$ with accounting self--action effects
within supersymmetry perturbation theory. A comparison of the
dependencies obtained with experimental data given in Section V
shows that the scheme developed allows to present in a
self--consistent manner main peculiarities of the microphase
separation in the homopolymer--oligomer systems with associative
coupling.

\section{Generic formalism}

The problem under consideration is addressed to the definition of
the effective law of motion $c(n)$ that determines a sequence of
oligomer alternation along the homopolymer chain by means of
specifying the occupation number being $c(n)=1$ if oligomer is
attached to the segment $n$, and $c(n)=0$ otherwise. When the
index of the homopolymer chain $N\to\infty$, the argument $n$ may
be considered as a continuum one, and we are ventured to start
with Euler equation \cite{10}
\begin{equation}
{\delta S\over\delta c}-{{\rm d}\over{\rm d}n}
{\delta S\over\delta {\dot c}}={\delta R\over\delta {\dot c}}\
\label{2}
\end{equation}
where dot denotes derivative with respect to the segment number
$n$, action $S$ and dissipative functional $R$ take the usual
forms
\begin{equation}
S\{c(n)\}\equiv\int\limits_0^N L\big(c(n),\dot c(n)\big){\rm d}n,\
R=\frac{\Theta}{2}\int\limits_0^N\big(\dot c(n)\big)^2{\rm d}n
\label{3}
\end{equation}
being defined by the Lagrangian $L\big(c(n),\dot c(n)\big)$ and
the damping coefficient $\Theta$, respectively. The total action
$S=K-\Pi$ is determined by a "kinetic" contribution $K$ of
inhomogeneity in the oligomers distribution and "potential"
component $\Pi\equiv V_0+V$ caused by the interaction between
homopolymer and oligomers
\begin{equation}
V_0=T\frac{\tau}{2}\int\limits_0^N \big(c(n)\big)^2{\rm d}n
\label{4}
\end{equation}
and self--action contribution
\begin{equation}
V=T\int\limits_{0}^{N}v\big(c(n)\big){\rm d}n,\quad
v\equiv\frac{\mu}{3!}c^3+\frac{\lambda}{4!}c^4.
\label{5}
\end{equation}
Here, $T$ is temperature measured in energy units, factor $\tau$
determines the strength of the hydrogen bonding, multipliers
$\mu$, $\lambda$ are self--action parameters.

In comparison with the above standard approach, the determination
of the contribution of inhomogeneity along polymeric chain is much
more delicate problem. Indeed, the bare magnitude can be written
in the form of the usual kinetic action
\begin{equation}
K=T\frac{m}{2}\int\limits_0^N \left (\frac{{\rm d} c}{{\rm
d}n}\right)^2 {\rm d}n
\label{6}
\end{equation}
where an effective mass $m$ appears as a temperature fluctuating
parameter with mean value $\bar m$ and variance 
$\overline{(m-\bar {m})^2}\equiv\sigma^2$ (bar denotes the
average, as usually). Then, after averaging exponent $\exp(-K/T)$
over the Gaussian distribution of the bare mass $m$, we obtain the
effective kinetic action in the following form:
\begin{eqnarray}
{\mathcal{K}}=\bar{K}+\tilde{K};\quad
\bar{K}\equiv T\frac{\bar m}{2}\int\limits_0^N
\big(\dot c(n)\big)^2{\rm d}n, \\
\tilde{K}\equiv -T\frac{\sigma^2}{8}
\int\limits_0^N\int\limits_0^N \big(\dot c(n)\big)^2\big(\dot
c(n')\big)^2{\rm d}n{\rm d}n'. \label{7}
\end{eqnarray}

As a result, total action takes the final form
\begin{widetext}
\begin{eqnarray}
S=T\frac{\bar m}{2}\int\limits_0^N\big(\dot c(n)\big)^2{\rm d}n -
T\frac{\sigma^2}{8}\int\limits_0^N\int\limits_0^N \big(\dot
c(n)\big)^2\big(\dot c(n')\big)^2{\rm d}n{\rm d}n'-
T\frac{\tau}{2}\int\limits_0^N\big(c(n)\big)^2{\rm d}n - V
\label{8}
\end{eqnarray}
\end{widetext}
where self--action potential $V$ is given by Eqs.
(\ref{5}). Respectively, Euler equation (\ref{2}) arrives at the
equation of effective motion
\begin{eqnarray}
\left({\bar m}-\tilde\Delta\right)\ddot c+ {\Theta\over T}\dot
c+\tau c=-v' \label{9}
\end{eqnarray}
where one notices
\begin{eqnarray}
v'\equiv T^{-1}{\delta V\{c(n)\}\over\delta c(n)}= \frac{\partial
v}{\partial c},\ {\tilde\Delta}\equiv\frac{\sigma^2}{2}
\int\limits_0^N\big(\dot c(n')\big)^2{\rm d}n'. \label{10}
\end{eqnarray}
By introducing the effective mass ${\tilde m}$, characteristic
number of correlating segments $n_c$ and $\delta$--correlated
source $\zeta(n)$ in accordance with definitions
\begin{eqnarray}
&{\tilde{m}}\equiv{\bar m}-\tilde\Delta,\quad
n_c\equiv{\Theta\over T},&\label{11a}\\
&\langle\zeta(n)\rangle=0,\quad
\left\langle\zeta(n)\zeta(n')\right\rangle=\delta(n-n'),&
\label{11}
\end{eqnarray}
one obtains Langevin equation of inertial type
\begin{equation}
{\tilde m}\ddot c+n_c\dot c=-(\tau c+v')+\zeta. \label{12}
\end{equation}

Making use of the field scheme \cite{10} allows to express the
noise $\zeta$ in terms of an additional degree of freedom $p$
being the momentum conjugated to the effective coordinate $c$.
Following this line, one has to introduce the generating
functional
\begin{equation}
Z\{c(n)\}\equiv\left<\prod_{n}\delta\left\{ {\tilde m}\ddot
c+n_c\dot{c}+\tau c+v'-\zeta\right\} {\rm
det}\left|{\delta\zeta\over\delta c}\right|\right> \label{13}
\end{equation}
being the average over the noise $\zeta(n)$ where
$\delta$--function accounts for the equation of motion (\ref{12}),
and the determinant is Jacobian of transition from $\zeta(n)$ to
$c(n)$ that is equal $\Theta/T\equiv n_c $. Then, making use of
the functional Laplace representation of $\delta$--function over a
ghost field $p(n)$ and averaging Eq.(\ref{13}) over Gaussian
distribution being related to Eqs.(\ref{11}), we derive to the
standard form \cite{10}
\begin{eqnarray}
Z\{c(n)\}=\int\exp\left[-{\mathcal S}\{c(n), p(n)\}\right]{\rm
D}p(n), 
\\ {\mathcal S}\{c(n),
p(n)\}\equiv\int{\mathcal L}\big(c(n), p(n)\big){\rm d}n \label{14}
\end{eqnarray}
where effective Lagrangian is introduced
\begin{equation}
{\mathcal L}= ({\tilde m}\ddot{c}+n_c\dot{c}+\tau c+v')(n_c^{-1}p)
-{1\over 2}(n_c^{-1}p)^2. \label{15}
\end{equation}
According to Euler equations
\begin{equation}
{\partial{\mathcal L}\over\partial x}-{{\rm d}\over{\rm d}n}
{\partial{\mathcal L}\over\partial{\dot x}}+ {{\rm d}^2\over{\rm
d}n^2} {\partial{\mathcal L}\over\partial{\ddot x}}=0,\quad
x\equiv \{c, p\} \label{16}
\end{equation}
effective motion in the phase space is determined by the system
\begin{eqnarray}
&{\tilde{m}}\ddot{c}+n_c\dot c=-(\tau {c}+v')+(n_c^{-1}p),&
\label{17} \\
&{\tilde{m}}\ddot{p}-n_c\dot p=-\left(\tau+v''\right)p.&
\label{18}
\end{eqnarray}
A comparison of the first of these equations with Eq. (\ref{12})
shows that the conjugated momentum $p$ appears as the most
probable value of the renormalized noise amplitude $n_c\zeta$
\cite{13}.

\section{Supersymmetry representation of correlation in oligomer distribution}

Eqs.(\ref{17}), (\ref{18}) represents a system of nonlinear
equations whose solution demands a use of the perturbation theory
with respect to the self-action parameters $\lambda$, $\mu$ and of
the self--consistency procedure to determine an effective mass
$\tilde{m}\{c(n)\}$. However, because we are interested in the
knowledge not of laws of motion $c(n)$ and $p(n)$, but only of the
frequency of oligomer alternation along the homopolymer chain, it
is appropriate to restrict ourself to an investigation of the
corresponding correlators. The latters reduce to autocorrelator
$S(n)=\langle c^2(n)\rangle$, and retarded and advanced Green
functions, $G_{-}(n)=\langle c(n)p(0)\rangle$ and
$G_{+}(n)=\langle p(n)c(0)\rangle$, respectively. As shows the
consideration given in \cite{11}~--~\cite{13}, it is convenient to
represent these correlators as the components of the
supercorrelator
\begin{equation}
C(z)\equiv\langle\phi^2(z)\rangle,\quad
z\equiv\{n,\vartheta\}
\label{19}
\end{equation}
of pseudovectors of the phase space
\begin{eqnarray}
\phi=c+(n_c^{-1}p)\vartheta
\label{20}
\end{eqnarray}
being built by making use of nilpotent variable $\vartheta$ which
satisfies the conditions:
\begin{eqnarray}
\vartheta^2=0,\quad \vartheta\vartheta'= \vartheta'\vartheta,\quad
\int{\rm d}\vartheta=0,\quad\int\vartheta ~{\rm d}\vartheta=1.
\label{21}
\end{eqnarray}
Along this line, the supercorrelator (\ref{19}) appears as a
pseudovector
\begin{equation}
{\bf C}=G_{+}{\bf A}+G_{-}{\bf B}+S{\bf T}
\label{22}
\end{equation}
spanned on set of the orts
\begin{equation}
{\bf A}(\vartheta,\vartheta')=\vartheta, \quad {\bf
B}(\vartheta,\vartheta')=\vartheta', \quad {\bf
T}(\vartheta,\vartheta')=1. \label{23}
\end{equation}
Introducing the functional product of some vectors $X$, $Y$, $Z$
in such a space:
\begin{equation}
X(\vartheta,\vartheta')=\int Y(\vartheta,\vartheta'')
Z(\vartheta'',\vartheta'){\rm d}\vartheta'',
\label{23a}
\end{equation}
it is easy to  see that orts (\ref{23}) are noncommutative to obey
to the multiplication rules given in Table I.
\begin{table}
\caption{\label{tab:table1}}
\begin{ruledtabular}
\begin{tabular}{|c| c| c| c|}
&&& \\
${l\backslash r}$ & ${\bf T}$
& ${\bf A}$ & ${\bf B}$\\
\hline &&&
\\
${\bf T}$ & $0$ & ${\bf T}$ & $0$ \\ \hline
&&& \\
${\bf A}$ & $0$ & ${\bf A}$ & $0$ \\ \hline
&&& \\
${\bf B}$ & ${\bf T}$ & $0$ & ${\bf B}$ \\ 
\end{tabular}
\end{ruledtabular}
\end{table}
\noindent As a result, making use of the supercorrelator
(\ref{19}) presents a big advantage in analytical calculations.

Under suppression of the inhomogeneity fluctuations along the
homopolymer chain ($\sigma=0$), the action (\ref{14}) with the
Lagrangian (\ref{15}) written within the harmonic approximation
($v(c)={\rm const}$) takes the canonical form
\begin{eqnarray}
{\mathcal S}_0\{\phi(z)\}=\frac{1}{2}
\int\phi(z)L(z)\phi(z){\rm d} z
\label{24}
\end{eqnarray}
with the linear operator
\begin{eqnarray}
L(z)=\tau(n)+D(z);\quad \tau(n)\equiv\tau+{\bar m}\partial^2_{n},
\label{25}\\
\partial^2_n=\frac{\partial^2}{\partial n^2},\quad
D(z)=-{\partial \over \partial \vartheta}+ n_c\left(1-2\vartheta
{\partial\over\partial\vartheta}\right){\partial\over\partial n}.
\label{25a}
\end{eqnarray}
This operator defines the bare supercorrelator according to the
relation
\begin{equation}
C^{(0)}(z)\equiv L^{-1}(z)\delta(\vartheta,\vartheta'),\quad
\delta(\vartheta,\vartheta')=\vartheta+\vartheta'.
\label{26}
\end{equation}
Taking into account condition $D^2=n_c^2\partial_n^2$ \cite{11},
one obtains
\begin{equation}
C^{(0)}=\frac{\left[\tau(n)-D\right]
\delta(\vartheta,\vartheta')}{\tau^2(n)-n_c^{2}\partial_n^2}.
\label{27}
\end{equation}
Using Fourier transformation over the frequency $\nu$, we arrive
to the expression
\begin{widetext}
\begin{eqnarray}
C^{(0)}=\frac{1+\left[\tau(\nu)-{\rm i}n_c\nu\right]\vartheta+
\left[\tau(\nu)+{\rm i}n_c\nu\right]\vartheta'}
{\tau^2(\nu)+n_c^{2}\nu^2},\quad \tau(\nu)\equiv \tau-{\bar
m}\nu^2. \label{28}
\end{eqnarray}
Then, taking into account Eqs. (\ref{22}), (\ref{23}), we get
standard equalities for the main correlators
\begin{equation}
G_{\pm}^{(0)}=\left[\tau(\nu)\pm{\rm i}n_c\nu\right]^{-1}, \quad
S^{(0)}\equiv G_{+}^{(0)}G_{-}^{(0)}=
\left[\tau^2(\nu)+n_c^{2}\nu^2\right]^{-1}. \label{29}
\end{equation}
\end{widetext}

An explicit form of linear operator
\begin{equation}
{\bf L}=L_{+}{\bf A}+L_{-}{\bf B}+L{\bf T}
\label{30}
\end{equation}
obeying to equality ${\bf L}\equiv \left({\bf
C}^{(0)}\right)^{-1}$ will be needed below. Using the equality
\cite{11}
\begin{equation}
{\bf C}^{-1}=G_+^{-1}{\bf A}+G_-^{-1}{\bf B}- G_+^{-1}S
G_-^{-1}{\bf T}, \label{31}
\end{equation}
we arrive easily to the components
\begin{equation}
L_{\pm}=\tau(\nu)\pm{\rm i}n_c\nu, \quad L=-1. \label{32}
\end{equation}

To proceed, let us consider the effective interaction term in
action (\ref{8})
\begin{equation}
\tilde{K}\equiv -T\frac{\sigma^2}{2} \iint\frac{{\rm d}\nu_1{\rm
d}\nu_2}{(2\pi)^2} \nu_1^2\nu_2^2|c(\nu_1)|^2|c(\nu_2)|^2
\label{7a}
\end{equation}
taken in the frequency representation. Within the mean--field
approximation, one has
\begin{widetext}
\begin{eqnarray}
|c(\nu_1)|^2|c(\nu_2)|^2\Rightarrow\left\langle|c(\nu_1)
|^2\right\rangle |c(\nu_2)|^2+|c(\nu_1)|^2\left\langle|c(\nu_2)|^2
\right\rangle
=S(\nu_1)|c(\nu_2)|^2+|c(\nu_1)|^2 S(\nu_2)
\label{33}
\end{eqnarray}
\end{widetext}
and the fluctuational component of the inhomogeneity action
(\ref{7a}) takes the form
\begin{eqnarray}
{\tilde{\mathcal K}}\{\phi\}= -T\Delta\int\frac{{\rm d}\nu}{2\pi}
\nu^2|\phi(\nu,\vartheta)|^2 \vartheta{\rm d}\vartheta \label{34}
\end{eqnarray}
where parameter ${\tilde\Delta}$ given by Eq. (\ref{10})
reduces to averaged magnitude
\begin{eqnarray}
\Delta=\sigma^2
\int\frac{{\rm d}\nu}{2\pi}\nu^2 S(\nu)
\Rightarrow\sigma^2\int\frac{{\rm d}\nu}{2\pi}\nu^2 C(\nu,\vartheta)
\vartheta{\rm d}\vartheta.
\label{35}
\end{eqnarray}
As a result, the bare mass $\bar m$ in the action $S_0$ given by
Eqs. (\ref{24}), (\ref{25}) is replaced by the effective quantity
\begin{eqnarray}
m_{ef}\equiv{\bar m}-\Delta \label{36}
\end{eqnarray}
being averaged value of the fluctuating mass (\ref{11a}).

To finish supersymmetry representation of the action (\ref{14})
defined by the Lagrangian (\ref{15}), one should add to
Eqs.(\ref{24}), (\ref{34}) the self--action term
\begin{eqnarray}
{\mathcal V}\{\phi(z)\}=
\int v\big(\phi(z)\big){\rm d}z,\quad
z\equiv\{n,\vartheta\}
\label{34a}
\end{eqnarray}
with the expansion (\ref{5}). Then, the standard perturbation
theory gives the symbolic expression \cite{10}
\begin{equation}
\begin{split}
& \Sigma(\vartheta_1,\vartheta_2,n) \\
& =\frac{\mu^2}{2!}\big(C(\vartheta_1,
\vartheta_2;n)\big)^2+\frac{\lambda^2}{3!}
\big(C(\vartheta_1,\vartheta_2;n)\big)^3 \label{40}
\end{split}
\end{equation}
for the self-energy function $\Sigma(\vartheta_1,\vartheta_2,n)$
defined by the following equation for the $n$-point dressed
supercorrelator
\begin{equation}
\begin{split}
& C^{(n)}(\vartheta,\vartheta') \\ & =\iint
C^{(0)}(\vartheta,\vartheta_1)\Sigma^{(n)}(\vartheta_1,\vartheta_2)C^{(0)}
(\vartheta_2,\vartheta'){\rm d}\vartheta_1{\rm d}\vartheta_2.
\label{40a}
\end{split}
\end{equation}
However, detailed analysis \cite{Kurchan} shows that the
multiplication rules given by Table 1 has to be replaced by the
rules of Table 2.
\begin{table}[!h]
\caption{\label{tab:table2}}
\begin{ruledtabular}
\begin{tabular}{|c|c|c|c|} 
&&&  \\ &$T (\vartheta,\vartheta')$ & $A(\vartheta,\vartheta')$ &
$B(\vartheta,\vartheta')$ \\ \hline
&&& \\
$T(\vartheta,\vartheta')$ & $T(\vartheta,\vartheta')$ &
$A(\vartheta,\vartheta')$ & $B(\vartheta,\vartheta')$\\ \hline
&&& \\
$A(\vartheta,\vartheta')$ & $ A(\vartheta,\vartheta')$ & $0$ &
$0$\\ \hline
&&& \\
$B(\vartheta,\vartheta')$ &
$B(\vartheta,\vartheta') $ & $0$ & $0 $\\ 
\end{tabular}
\end{ruledtabular}
\end{table}
Then, the components of the pseudovector
\begin{equation}
{\bf \Sigma}=\Sigma_{+}{\bf A}+\Sigma_{-}{\bf B}+\Sigma{\bf T}
\label{38}
\end{equation}
take the following forms:
\begin{widetext}
\begin{eqnarray}
\Sigma(\nu)=\frac{\mu^2}{2}\int\frac{{\rm d}\nu_1}{2\pi} S(\nu_1)
S(\nu-\nu_1)+\frac{\lambda^2}{6}\iint \frac{{\rm d}\nu_1 {\rm
d}\nu_2}{(2\pi)^2} S(\nu_1)S(\nu_2)S(\nu-\nu_1-\nu_2),
\label{41} \\
\Sigma_{\pm}(\nu)=\mu^2\int\frac{{\rm d}\nu_1}{2\pi} S(\nu_1)
G_{\pm}(\nu-\nu_1)+ \frac{\lambda^2}{2}\iint\frac{{\rm d}\nu_1
{\rm d}\nu_2}{(2\pi)^2} S(\nu_1)S(\nu_2)G_{\pm}(\nu-\nu_1-\nu_2).
\label{42}
\end{eqnarray}
Making use of the theory of residues (see Appendix B) with the
correlators (\ref{29}), where the frequency dependent parameter
$\tau(\nu)\equiv \tau-{\bar m}\nu^2$ is replaced by the bare one
$\tau$, arrives at the equalities (\ref{a11}), (\ref{a11a}) which
take the form
\begin{eqnarray}
\Sigma\simeq (8\tau^3
n_c)^{-1}\left[\left(\mu^2+\frac{\lambda^2}{3^2\tau n_c}\right)-
\left(\frac{\mu^2}{2^2}+ \frac{\lambda^2}{3^4\tau n_c }\right)\xi
^2\right],&
\label{43} \\
\Sigma_{\pm}\simeq (4\tau^2 n_c
)^{-1}\left[\left(\mu^2+\frac{\lambda^2}{6\tau n_c}\right)
\mp{{\rm i}\over 2}\left(\mu^2+\frac{\lambda^2}{3^2\tau
n_c}\right)\xi  -\left(\frac{\mu^2}{2^2}+ \frac{\lambda^2}{2\cdot
3^3\tau n_c}\right)\xi ^2\right] \label{44}
\end{eqnarray}
\end{widetext}
within the hydrodynamic limit $\xi \equiv \nu/\omega_s\ll 1$,
$\omega_s\equiv\tau/n_c$.

Self--consistent behavior of the system under consideration is
described by the generalized Dyson equation \cite{11}
\begin{equation}
{\bf C}^{-1}={\bf L}-{\bf\Sigma}. \label{37}
\end{equation}
In the component representation this equality arrives at the
equations
\begin{eqnarray}
&&S=(\Sigma-L)G_{+}G_{-},\label{39a}\\
&&G^{-1}_{\pm}=L_{\pm}-\Sigma_{\pm}.
\label{39}
\end{eqnarray}
Combination of Eqs. (\ref{32}), (\ref{43}), (\ref{44}) arrives at
the final equations for main correlators within hydrodynamical
limit $\xi \ll 1$:
\begin{equation}
\begin{split}
G_{\pm}^{-1}=\left[\tau-(4\tau^2 n_c)^{-1}\left(\mu^2+
\frac{\lambda^2}{6\tau n_c}\right)\right] \\ 
\pm\ {\rm i}\left[\tau+ (8\tau^2 n_c)^{-1}\left(\mu^2+
\frac{\lambda^2}{3^2\tau n_c}\right)\right]\xi \\ 
-\left[\frac{\tau^2 m_{ef}}{n_c^2}-(4\tau^2 n_c
)^{-1}\left(\frac{\mu^2}{2^2}+ \frac{\lambda^2}{2\cdot 3^3\tau n_c
}\right)\right]\xi ^2. \label{45}
\end{split}
\end{equation}

\section{Determination of the period of microphase structure}

Our consideration is based on the obvious equality for the long
space period $L=2l+D$ where $l$ is the oligomer chain length, $D$
is the thickness of the homopolymer layer being fixed by the
inverse share $X^{-1}$ of average number of the hydrogen bonds
(see Figure \ref{structure}). Physically, this value is reduced to
the magnitude $2\pi/\omega$ determined by the circular frequency
$\omega$ in the alternation of the oligomer heads along the
homopolymer chain. Then, the long space period is expressed by the
following equality \cite{2} (see Appendix A)
\begin{equation}
L=2l+D_0\omega^{-1},\quad
D_0\equiv\left(2\pi\chi^{1/6}n^{-1/3}\right)b\geq b \label{1}
\end{equation}
where $\chi\leq 10^{-1}$ is the Flory parameter, $n\sim 10$ is the
number of segments in oligomer chain, $b$ is the segment length.

To obtain the frequency $\omega$, one has to determine firstly the
effective mass $m_{ef}$ given by Eqs. (\ref{36}), (\ref{35}).
Usage of the theory of residues (see Appendix B) with the
structure factor (\ref{39a}) and Green function (\ref{45}) arrives
at the renormalization mass parameter
\begin{eqnarray}
\Delta=\frac{\sigma^2}{2 m_{ef}n_c} \left(1+{1\over
2^5}~\frac{n_c\mu^2}{\tau^4 m_{ef}}+ \frac{1}{2^3\cdot
3^4}~\frac{\lambda^2}{\tau^5 m_{ef}}\right) \label{46a}
\end{eqnarray}
where only the terms of the second order of smallness over the
parameters $\mu, \lambda$ of the self--action (\ref{5}) are kept.
Inserting here Eqs. (\ref{11a}), (\ref{36}), we obtain the
equations for determination of the effective mass as a function of
the temperature:
\begin{eqnarray}
&m_{ef}=\mu\bar{m},\quad\mu=\mu(T);& \label{47} \\
&4\mu(1-\mu)=\frac{T}{
T_{c0}}\left(1+\frac{\alpha+\beta/T}{\mu}\right)& \label{48}
\end{eqnarray}
where
\begin{eqnarray}
T_{c0}\equiv\left(\frac{\bar{m}}{\sigma}\right)^2\frac{\Theta}{2},\quad
\alpha\equiv\frac{1}{2^3\cdot
3^4}~\frac{\lambda^2}{\tau^5\bar{m}},~\beta\equiv{1\over
2^5}~\frac{\Theta\mu^2}{\tau^4\bar{m}}.\label{49}
\end{eqnarray}
Numerical solution of Eq. (\ref{48}) for different values of
$\alpha$ and $\beta$ allows us to estimate an influence of the
self--action on the effective mass. It turned out that even small
variation of the parameter $\alpha$ substantially changes the
shape of the dependence $\mu(T)$, whereas the parameter $\beta$
almost does not affect it, and we can put $\beta=0$ for the sake
of simplicity. This means physically that the cubic anharmonicity
in the self--action potential energy (\ref{5}) is irrelevant to
the microphase separation phenomenon.

\begin{figure}[!h]
\epsfig{file=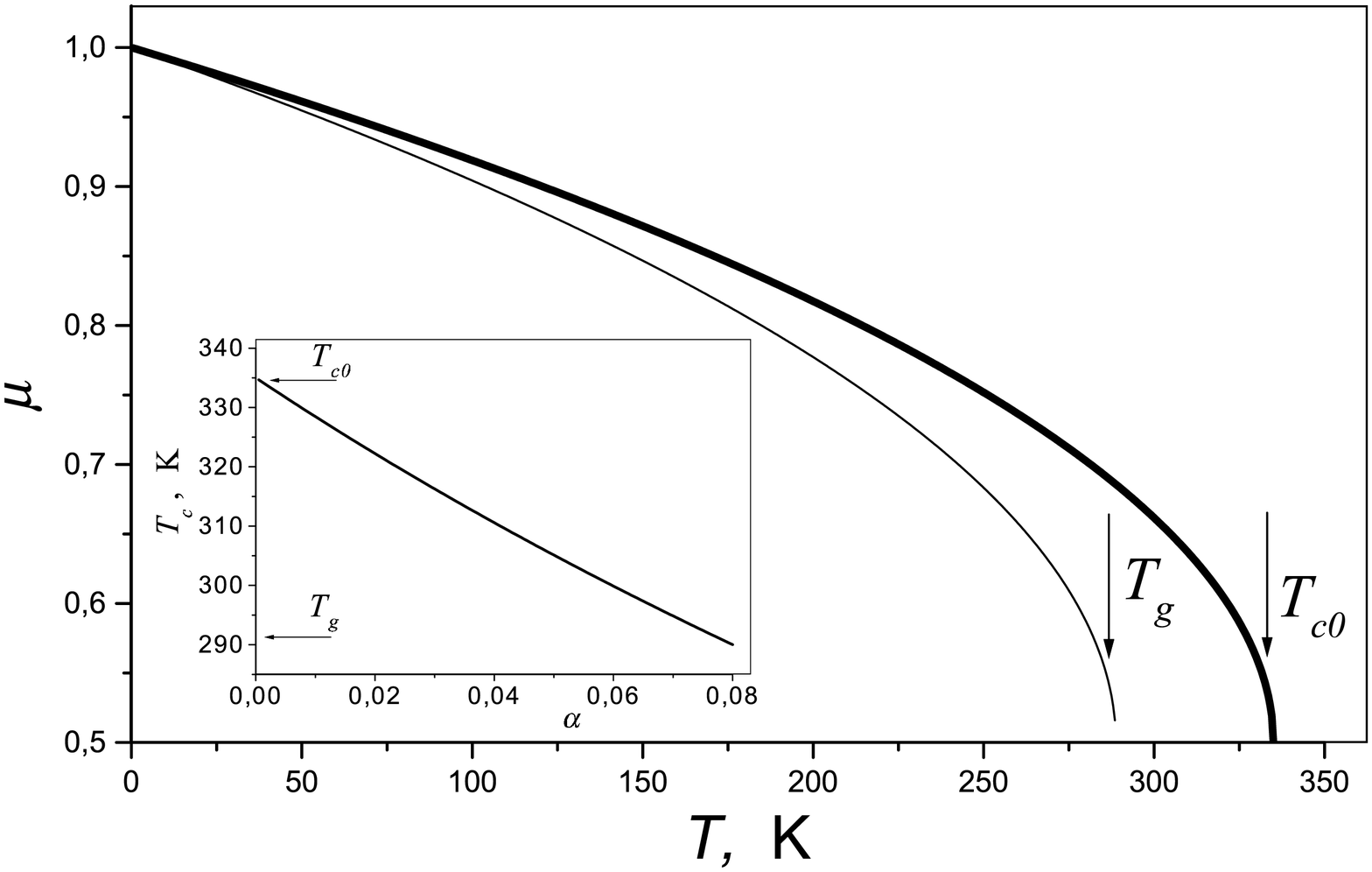,width=80mm,bbllx=18mm,bblly=5mm,bburx=287mm,bbury=180mm,angle=0}
\caption{Temperature dependence of the inhomogeneity parameter
$\mu$ for different values of the self--action parameter $\alpha$:
thick curve relates to $\alpha=0$, thin one --- to $\alpha=0.08$
($T_{c0}=337{\rm K}$). Inset: the  temperature of order--disorder
transition $T_c$ versus the parameter $\alpha$.} \label{mudep}
\end{figure}

The smallness of the self--action parameters $\alpha$, $\beta$
allows us to solve Eq. (\ref{48}) analytically. In so doing, one
has to replace the required dependence $\mu(T)$ in the right hand
part of Eq. (\ref{48}) by the bare dependence
\begin{eqnarray}
\mu_0(T)=\frac{1}{2}\left(1+\sqrt{1-\frac{T}{T_{c0}}}\right),
\label{50a}
\end{eqnarray}
that is a solution of this equation at $\alpha=\beta=0$. As a
result, we obtain the simple dependence
\begin{eqnarray}
\mu(T)=\frac{1}{2}\left(1+\sqrt{1-\frac{T}{T_c}}\right)
\label{50}
\end{eqnarray}
with a characteristic temperature
\begin{eqnarray}
T_c\equiv T_{c0}\left(1-2\alpha\right) \label{51}
\end{eqnarray}
where the scale $T_{c0}$ is given by the first of Eqs. (\ref{49})
(the multiplier should be put $\mu_0(T_{c0})\simeq 1/2$ due to the
smallness of the parameter $\alpha\ll 1$). According to Eq.
(\ref{50}), with the temperature increasing the effective mass
(\ref{47}) decreases monotonously from the bare magnitude
$\bar{m}$ at $T=0$ to $\bar{m}/2$ at $T=T_c$ (see the main panel
in Figure \ref{mudep}). The critical temperature $T_c$ determines
the point of the order-disorder transition according to the
condition
$$\frac{{\rm d}\mu}{{\rm d}T}=-\infty.$$
Resulting dependence $T_{c}$ on the self--action parameter
$\alpha$ is shown in the inset of Figure \ref{mudep}. It is
principally important that the bigger value of the self--action
parameter $\alpha$, the more narrow is the temperature domain
$T_g\div T_c$ where the microphase separated structure is possible
($T_g$ being a glassing temperature). In other words,  the
self--action effect leads to the shrinking the region of the
ordered structure because the critical temperature $T_c$ reaches
the boundary magnitude $T_g$ with increasing of $\alpha$ before
the magnitude $\alpha\simeq 0.08$.

The divergency condition $G_-^{-1}=0$ of the Green function
(\ref{45}) gives the proper frequency
\begin{eqnarray}
\nu_0=\pm\omega-{\rm i}\varpi,\quad
\omega\equiv\sqrt{\omega_0^2-\varpi^2} \label{53}
\end{eqnarray}
of the oligomer alternation along the homopolymer chain. Real and
imaginary parts are determined by the expressions
\begin{equation}\label{fr}
\begin{split}
\omega_0\equiv\frac{{\bar\omega}_0}{\sqrt{\mu(T)}}\left[1+3\alpha\left(1-\frac{9}{8}~\frac{T^2}{T_0^2}\right)\right],\\
\varpi\equiv\frac{{\bar\omega}_0}{\mu(T)}~\frac{T_0}{T}\left[1+6\alpha\left(1+\frac{3}{8}~\frac{T^2}{T_0^2}\right)\right]
\end{split}
\end{equation}
where the dependence $\mu(T)$ is defined by Eqs. (\ref{50}),
(\ref{51}); the effective mass in parenthesis after the factor
$\alpha\ll 1$ is put to be equal to the value ${\bar m}/2$ related
to the critical temperature $T_c$; characteristic scales of both
frequency and temperature are introduced as follows:
\begin{equation}
{\bar\omega}_0\equiv\sqrt{\frac{\tau}{{\bar m}}},\quad
T_0\equiv\frac{\Theta}{2\sqrt{{\bar m}\tau}}.
\end{equation}
As a result, combination of Eqs.(\ref{1}), (\ref{53}) and
(\ref{fr}) leads to the final result for the long space period
\begin{widetext}
\begin{equation}\label{1p}
L=2l+\frac{\mu(T)}{\sqrt{\mu(T)-(T_0/T)^2}}\left[1+\frac{3}{2}\alpha~\frac{1+8(T_0/T)^2+
\frac{9}{4}(T_0/T)^{-2}}{1-2(T_0/T)^2}\right]L_0
\end{equation}
\end{widetext}
where the characteristic length $L_0\equiv D_0/{\bar\omega}_0\sim
\sqrt{{\bar m}/\tau}~\chi^{1/6}b \propto \chi^{1/6}N^{1/2}$ is the
function of both parameters $\chi$ and $N$ being thermodynamically
independent. Thus, the first of the exponents in the scaling
relation $L_0\propto\chi^{a} N^{b}$ takes the magnitude $a=1/6$
inherent to the strong segregation regime, whereas the second one
($b=1/2$) is the same for the weak one \cite{a}. Note that the
obtained $\chi$--dependence is caused by the multiplier
$\chi^{1/6}$ in the generic relation (\ref{1}) that is relevant to
the former of above regimes, while the method developed addresses
to the latter.

\section{Discussions}

The behavior of the system under consideration is controlled by
the parameters $\bar{m}$, $\tau$ and $\sigma$ which determine the
temperature $T_c$ of the order--disorder transition and the long
space period $L$ given by Eqs. (\ref{51}), (\ref{1p}),
respectively. Moreover, there is the self--action parameter
$0<\alpha\ll 1$ whose value is limited by the magnitude
$\alpha_{max}\simeq 0.08$ (see insert in Figure \ref{mudep}). To
guarantee positive values of the radicand in Eq. (\ref{1p}) at the
critical temperature $T_c$, the above parameters have to be
constrained by the condition
\begin{equation}
\kappa\geq\sqrt{2} \label{cc}
\end{equation}
limiting magnitudes of the principal parameter
\begin{equation}
\kappa\equiv\frac{T_c}{T_0}=\sqrt{\tau{\bar m}}\left(\frac{{\bar m
}}{\sigma}\right)^2(1-2\alpha). \label{kappa}
\end{equation}
The minimal magnitude of $\kappa$ fixes the choice of the theory
parameters according to the condition
\begin{equation}
\sigma\leq 2^{-1/4}{\bar m}^{5/4}\tau^{1/4}(1-\alpha).
\label{1kap}
\end{equation}
It would seem from Eqs. (\ref{cc}), (\ref{kappa}) that the
decrease of the critical temperature $T_c$ with passing from the
ionically bonded system (such as P4VP--(DBSA)$_x$) to the hydrogen
bonded one (e.g., P4VP--(PDP)$_x$) is caused only by the growth of
the fluctuation parameter $\sigma$ with respect to the mean
magnitude of the inhomogeneity parameter $\bar m$. It appears,
however, that the main reason for such behavior is given by the
decrease of the mean--geometrical magnitude $\sqrt{{\bar m}\tau}$
of the principal coefficients in the generic Lagrangian (\ref{8})
(see below).

\begin{figure}[!h]
\epsfig{file=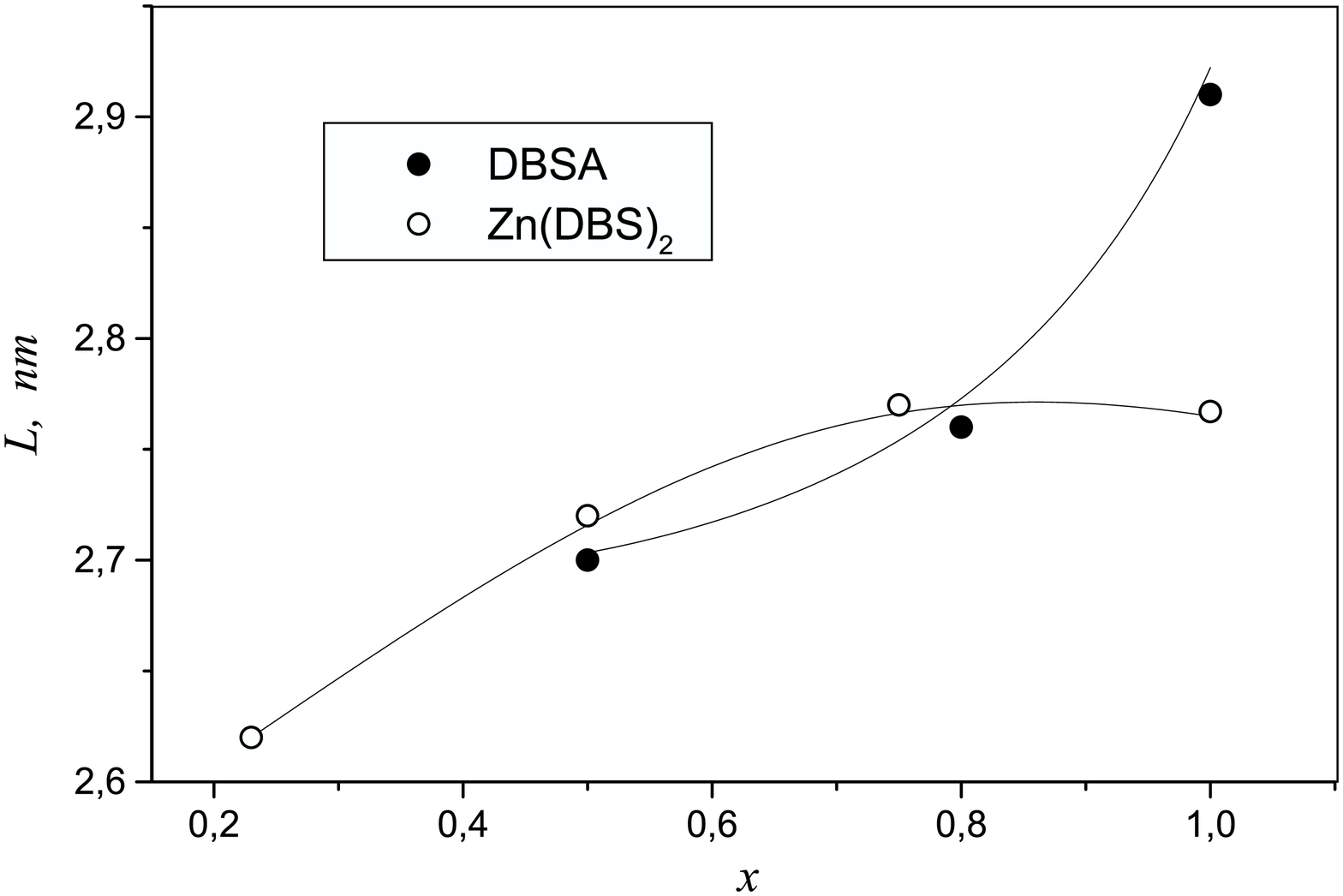,width=80mm,bbllx=10mm,bblly=10mm,bburx=275mm,bbury=187mm,angle=0}
\caption{Long space period in the strongly bonded systems as a
function of  the oligomeric fraction $x$. Solid lines represents
the results of fitting in accordance with Eq. (\ref{1p}).
Experimental data for  P4VP--(DBSA)$_x$ ($\bullet$) and
P4VP--(Zn(DBS)$_2$)$_x$ ($\circ$) at room temperature are taken
from Ref. \protect\cite{2}.} \label{Lxi}
\end{figure}

\begin{figure}[!h]
\epsfig{file=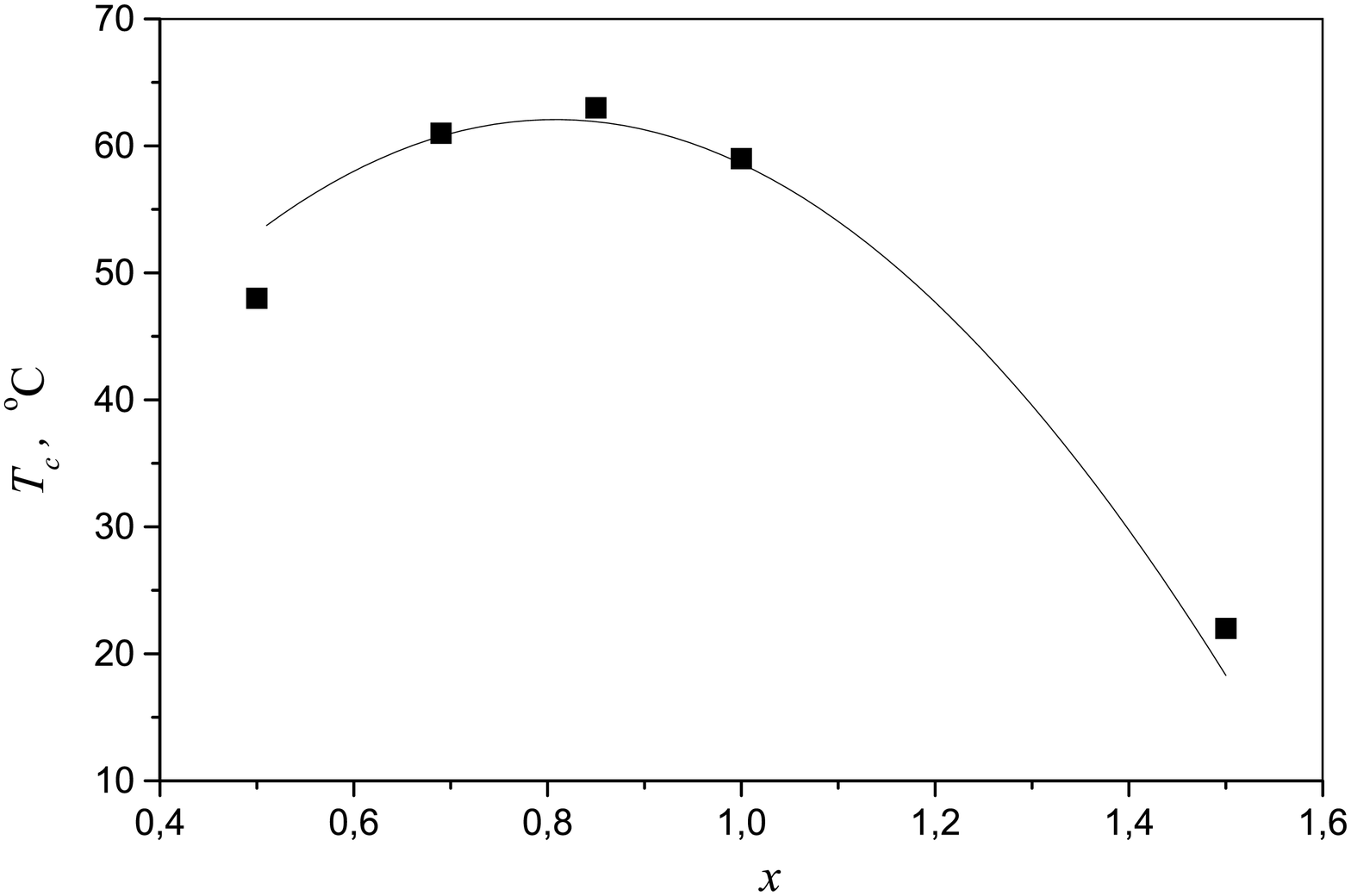,width=80mm,bbllx=13mm,bblly=11mm,bburx=275mm,bbury=187mm,angle=0}
\caption{Order-disorder transition temperature $T_c$  for the
weakly bonded system as a function of the oligomeric fraction $x$.
The solid line represents the dependence obtained by fitting
according to Eq. (\ref{51}). Experimental data for P4VP--(PDP)$_x$
($\blacksquare$)  are taken from Ref. \protect\cite{8}.}
\label{odt}
\end{figure}

\begin{figure}[!h]
\epsfig{file=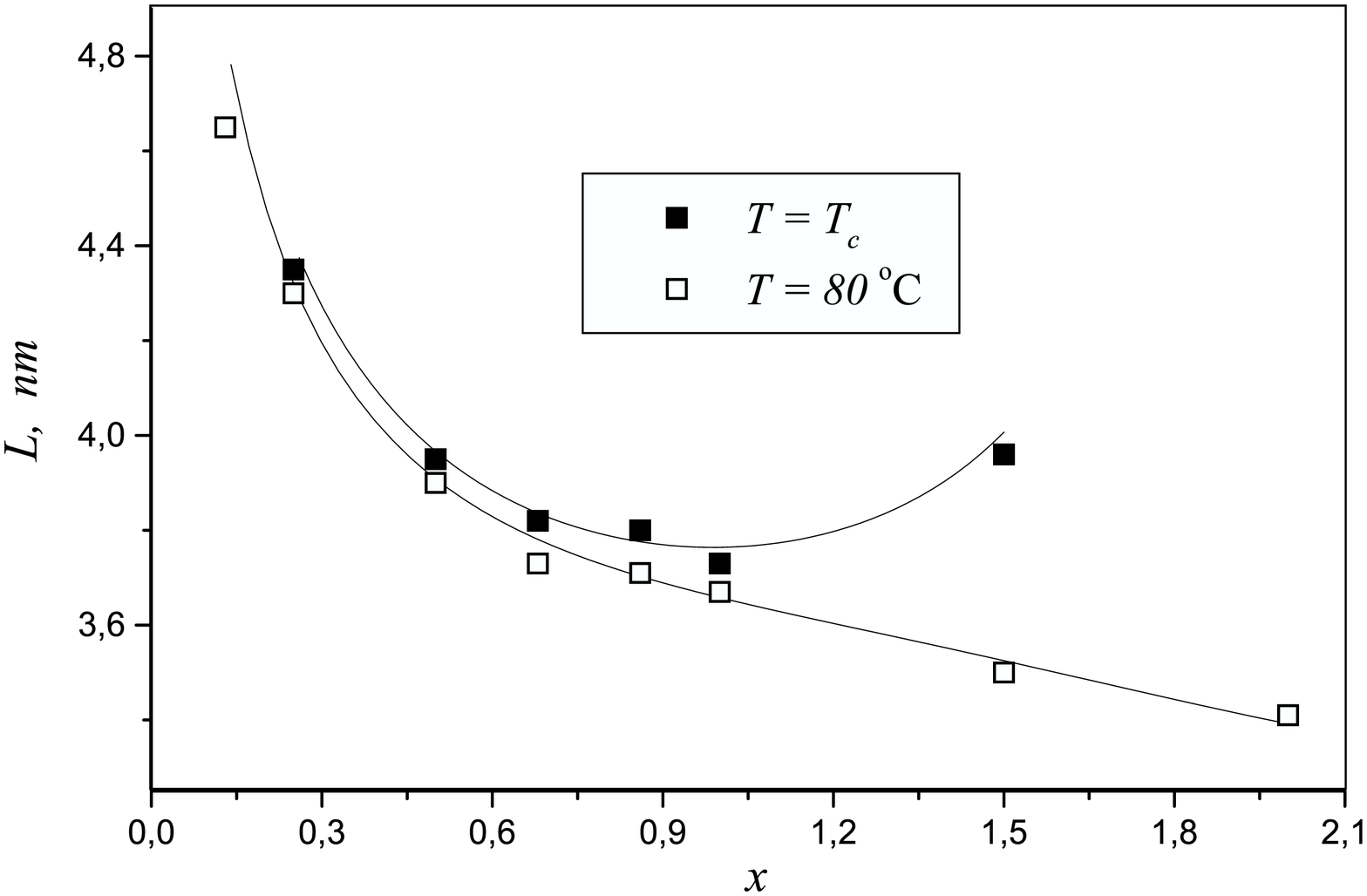,width=80mm,bbllx=13mm,bblly=11mm,bburx=275mm,bbury=187mm,angle=0}
\caption{Long space period in the weakly bonded systems as a
function of  the oligomeric fraction $x$. Solid lines represents
the results of fitting in accordance with Eq. (\ref{1p}).
Experimental data for P4VP--(PDP)$_x$ at temperature of
order--disorder transition $T_c$ ($\blacksquare$) and at
temperature $T=80^o {\rm C}$ ($\square$) are taken from Ref.
\protect\cite{8}.} \label{Lxh}
\end{figure}

\begin{figure}[!h]
\epsfig{file=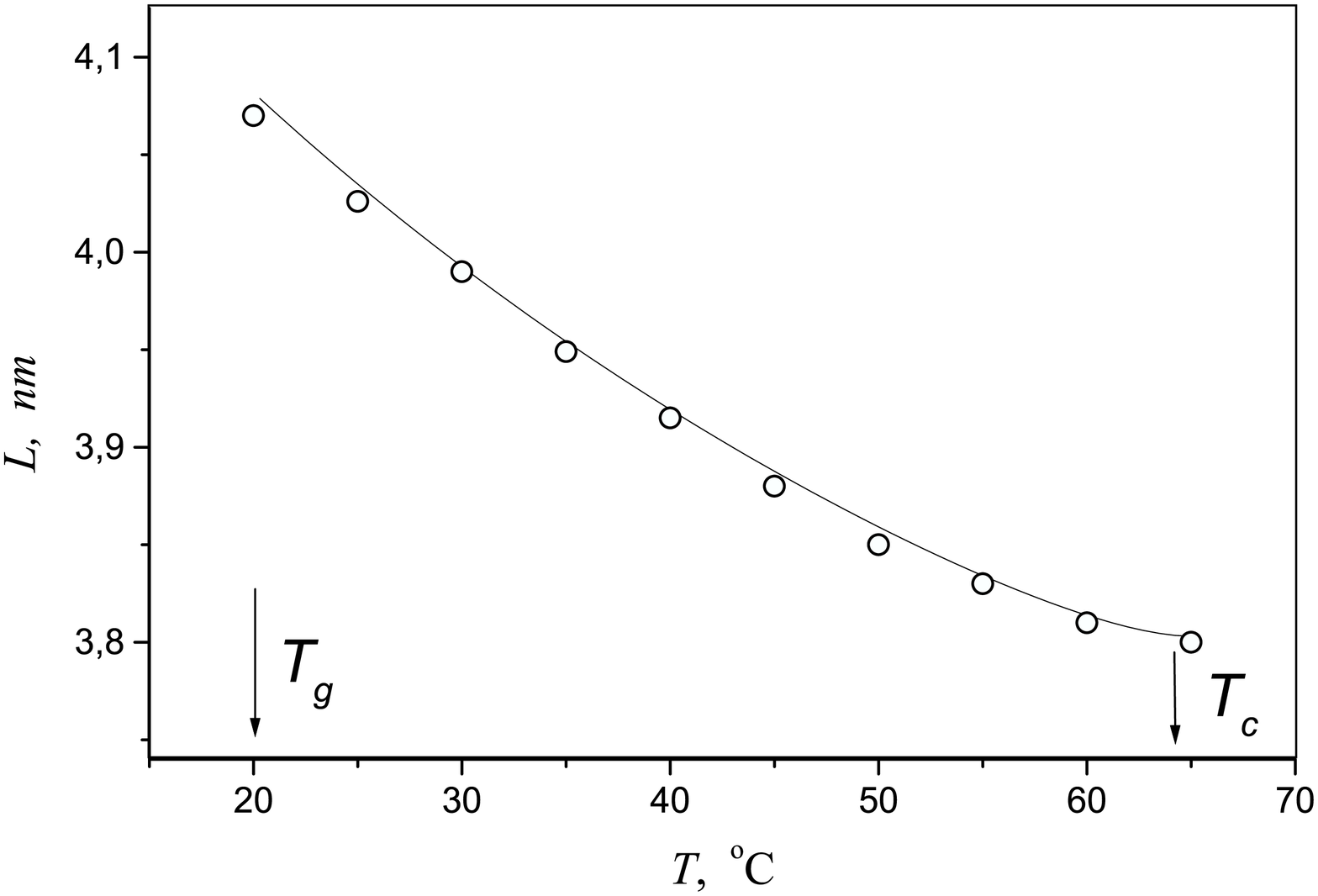,width=80mm,bbllx=11mm,bblly=4mm,bburx=275mm,bbury=187mm,angle=0}
\caption{Temperature dependence of the long space period in the
weakly bonded system. The solid line represents the dependence
obtained by fitting according to Eq. (\ref{1p}).  Experimental
data for  P4VP--(PDP)$_x$ at $x=0.85$ ($\circ$) are taken from
Ref. \protect\cite{8}.} \label{Lt}
\end{figure}

To clarify this problem and find explicit form of the dependencies
of the temperature of order-disorder transition $T_c$ and the
period $L$ on the oligomeric fraction $x$, we assume for main
theory parameters $\bar{m}$ and $\tau$ the three--parametric
relations:
\begin{equation}
\bar{m}=m_0+Ax(x_m-x),\quad \tau=\tau_0+Bx(x_{\tau}-x) \label{61}
\end{equation}
with positive constants $m_0$, $\tau_0$, $A$, $B$, $x_m$,
$x_{\tau}$ to be determined. Then, the fitting of the experimental
results shown in the Figure \ref{Lxi} in accordance with Eq.
(\ref{1p}) where $\tau$, $\bar{m}$ are given by Eq. (\ref{61})
leads to the following results for the ionically bonded systems:
\begin{itemize}
\item the mixtures P4VP--(DBSA)$_x ~~\qquad\qquad\qquad\qquad$
\break
\begin{equation}
\begin{split}
& m_0=18,\quad A=8,\quad  x_m=1.5;\\
& \tau_0=0.6,\quad  B=1.5,\quad  x_{\tau}=1.0;\\
& \alpha=0.01;\quad  b=1 nm;\quad  l=10 nm; \label{5a}
\end{split}
\end{equation}
\item the mixtures P4VP--(Zn(DBS)$_2$)$_x
~~\qquad\qquad\quad\quad$ \break
\begin{equation}
\begin{split}
& m_0=5.3,\quad  A=26,\quad  x_m=1.6;\\
& \tau_0=0.8,\quad  B=0.1,\quad  x_{\tau}=1.0;\\
&\alpha=0.01;\quad  b=1 nm;\quad l=10 nm. \label{5b}
\end{split}
\end{equation}
\end{itemize}
At $x=1$ one obtains ${\bar m}=22$, $\tau=0.6$ for
P4VP--(DBSA)$_x$ and ${\bar m}=20.9$, $\tau=0.8$ for
P4VP--(Zn(DBS)$_2$)$_x$.  Then, Eq. (\ref{kappa}) gives values
$\kappa=10^3$, $10^2$ at $\sigma= 1.31$, $4.18$, respectively.

Much more complicated situation occurs in the weakly bonded system
P4VP--(PDP)$_x$. Here, decrease of the parameter (\ref{kappa})
results in the narrowing of the temperature domain $T_0\div T_c$
of the phase separation. All parameters for this class of systems
can be determined by the combined fitting of a series of
experimental data for the critical temperature $T_c$ and the long
space period $L$ (see Figures \ref{odt} --- \ref{Lt}). First
constraints follow from the comparison of experimental points for
the temperature $T_c$ of order--disorder transition (see Figure
\ref{odt}) with fitting results based on Eq. (\ref{51}) at
$\alpha=0.01$, $l=10$ nm:
\begin{equation}
\frac{\Theta m_0^2}{\sigma^2}=562,\quad \frac{A}{m_0}=0.155,\quad
x_m=1.615. \label{5c}
\end{equation}
The following of parameters gives application of Eq. (\ref{1p})
for the long space period at the temperature $T=T_c$ to the data
shown in Figure \ref{Lxh} as the non-monotonous curve:
\begin{equation}
\frac{m_0}{\tau_0}=1499,\quad \frac{B}{\tau_0}=7.968,\quad
x_{\tau}=1.926. \label{5d}
\end{equation}
Finally, making use of the expression (\ref{1p}) and experimental
data for the temperature dependence of the long space period given
in Figure \ref{Lt} yields the last constraint
\begin{equation}
\frac{\Theta}{\sqrt{m_0\tau_0}}=924. \label{5e}
\end{equation}
As a result, taking $m_0=1$ at $x=1$ the magnitudes $A=0.155$,
$\Theta=23.87$, $\sigma=0.206$ are obtained to provide extremely
small value $\tau=5.6\cdot 10^{-3}$ of the hydrogen bonding
strength and the temperature scale $T_0=160$. At $\alpha=0.01$
this arrives to the rest of parameters $\kappa=2.07$,
$\lambda=6.22\cdot 10^{-6}$.

It is worthwhile to discuss separately the dependence of the long
space period on the oligomer/monomer ratio at the temperature
$T=80^0{\rm C}$ that relates to the monotonous decaying curve
shown in  Figure \ref{Lxh}. Because the maximal temperature of the
order--disorder transition is $T_c\approx 65^0{\rm C}$ to be
corresponded to $x=0.85$ (see Figure \ref{odt}), experimental data
related to $T=80^0{\rm C}$ are obtained for the temperature being
beyond of the region of the ordered structure ($T>T_c$). From the
physical point of view, at the critical temperature $T=T_c$ the
periodicity of the microphase separated structure formed is caused
by long--range correlations, whereas at $T=80^0{\rm C}$ only
short--range correlations hold to be determined by the homopolymer
backbone together with the hydrogen bound surfactant molecules
\cite{8}. Fitting of the experimental points for the dependence
$L(x)$ at the temperature $T=80^0{\rm C}$ can be done on the base
of Eq. (\ref{1p}) where one puts $\mu(T)=\mu(T_c)=1/2$. Then, the
values of the parameters obtained differ from those obtained for
$T=T_c$ by the following constraints:
\begin{equation}
\frac{B}{\tau_0}=4.2,\quad
x_{\tau}=3.27,\quad\frac{\Theta}{\sqrt{m_0\tau_0}}=543. \label{5f}
\end{equation}
Obviously, this difference is due to the temperature dependence of
the hydrogen bonding parameter $\tau$ in the potential energy
(\ref{4}).

To conclude our estimations, we notice the model developed
explains successfully a vast variety of peculiarities obtained
experimentally for various classes of homopolymer--oligomer
mixtures with the interactions of different strength. The resulted
data for strong, intermediate and weak coupled systems
P4VP--(DBSA)$_x$, P4VP-(Zn(DBS)$_2$)$_x$ and P4VP--(PDP)$_x$,
respectively, are given in Table III.
\begin{table}
\caption{\label{tab:table3}}
\begin{ruledtabular}
\begin{tabular}{|c|c|c|c|}
&\scriptsize{P4VP--(DBSA)$_x$} &
\scriptsize{P4VP--(Zn(DBS)$_2$)$_x$}&\scriptsize{P4VP--(PDP)$_x$}
\\
\hline
$m_0$ & $18$ & $5.3$ & $1.000$ \\ \hline
$A$ & $8$ & $26$ & $0.155$ \\ \hline
$x_m$ & $1.5$ & $1.6$ & $1.615$ \\ \hline
${\bar m}$ & $22$ & $20.9$ & $1.095$ \\ \hline
$\tau_0$ & $0.6$ & $0.8$ & $6.67\cdot 10^{-4}$ \\ \hline
$B$ & $1.5$ & $0.1$ & $5.3\cdot 10^{-3}$ \\ \hline
$x_{\tau}$ & $1.0$ & $1.0$ & $1.926$ \\ \hline
$\tau$ & $0.6$ & $0.8$ & $5.6\cdot 10^{-3}$ \\ \hline
$\sigma$ & $1.31$ & $4.18$ & $0.206$ \\ \hline
$\alpha$ & $0.01$ & $0.01$ & $0.01$ \\ \hline
$\lambda$ & $3.33$ & $6.66$ & $6.22\cdot 10^{-6}$ \\ \hline
$\Theta,{\rm K}$ & $10^3$ & $10^3$ & $23.87$ \\ \hline
$T_{c0},{\rm K}$ & $1.41\cdot 10^5$ & $1.25\cdot 10^4$ & $337$ \\ \hline
$T_c,{\rm K}$ & $1.38\cdot 10^5$ & $1.22\cdot 10^4$ & $331$ \\ \hline
$T_0,{\rm K}$ & $138$ & $122$ & $160$ \\ \hline
$\kappa$ & $10^3$ & $10^2$ & $2.07$ \\ \hline
$l, {\rm nm}$ & $10$ & $10$ & $10$ \\ \hline
$b, {\rm nm}$ & $1$ & $1$ & $1$ \\ 
\end{tabular}
\end{ruledtabular}
\end{table}
It is seen that the coupling weakening arrives to a decrease of
both inhomogeneity parameters $\bar{m}$ and $\sigma$, as well as
to the crucial decrease of the hydrogen bonding parameter $\tau$
and the self--action parameter $\lambda$, on the one hand, and the
characteristic temperatures $T_c$ and $\Theta$, on the other hand.
According to the relations (\ref{kappa}) this leads to extremely
large suppression of the value of the parameter $\kappa$ that
causes the crucial shrinking the temperature interval of the
microphase separation. An analogous effect is caused by the
self--action increase.

To get rid of a misunderstanding, we would like to stress a
composite character of the approach used. As it is mentioned in
Introduction, this circumstance is expressed by dividing the total
free energy (\ref{000}) into two terms, the first one $F_{ho}$ is
relevant to the non--associated homopolymer--oligomer mixture, the
second one $F_{hb}$ is caused by the hydrogen bonding. These terms
are caused by the interactions of principally different physical
nature: the behavior of the mixture of non--associated
homopolymers and oligomers is determined by the Flory parameter
$\chi$, characterizing unfavorable interactions between the the
oligomers and the rest of the system; the temperature induced
distribution of hydrogen bonds is determined by the parameter
$\tau$, giving the strength of this bonding. From the formal point
of view, both of the above contributions $F_{ho}(\chi,\phi)$ and
$F_{hb}(\tau,x)$ should have a similar dependencies on the state
parameters being (apart from the temperature) the volume fraction
of the homopolymer $\phi$ for the first  contribution, and the
oligomer/monomer ratio $x$ for the second one. Because the term
$F_{ho}\sim\chi\phi(1-\phi)$ involves the parabolic dependence on
the parameter $\phi$ bounded by maximal value $\phi=1$, we took
generalized parabolic approximation (\ref{61}) for the dependence
of the hydrogen bonding strength $\tau$ on the oligomer/monomer
ratio $x$ which may take values $x>1$.

Apart from the above difference in nature of the interactions, one
needs to emphasize at once the difference in the approaches used:
the mixture of non--associated homopolymers and oligomers had been
studied within the strong segregation limit \cite{2}, whereas for
the consideration of the hydrogen bonding we use opposite
approach. This difference is kept if the Flory parameter takes
large values $\chi\leq 10^{-1}$, whereas the hydrogen bonding
strength is relatively small ($\tau\ll 1$). Indeed, the formula
(\ref{1}) for the long space period was obtained within
approximation of the sharp interface, which thickness is
$\Delta\sim\chi^{-1/2}b\geq 3b$ to be relevant to the strong
segregation regime \cite{2}. In the consideration presented, we
have focused mainly on the study of the hydrogen bonding on the
base of the action (\ref{8}) that has the form of series in powers
of the order parameter $c$ and its derivatives ${\dot c}$. Such an
expansion supposes making use of the weak segregation limit
corresponding to the small values of the parameters $\bar m$ and
$\tau$.

Finally, it is worthwhile to discuss a difference with an usual
picture of the phase transitions that is caused by the
self--consistency condition (\ref{35}). A critical value of the
Flory parameter $\chi_c$ in usual copolymers is known to be caused
by the self--action effects. The accounting of these effects
arrives to replacement the bare parameter $\chi$ by the
renormalized value $\chi-\chi_c$ \cite{19}. However, in our case
the value of Flory parameter is so large that the temperature of
the separation of non--associated polymer--oligomer mixture is
negligible small. As a result, the role of $\chi$ passes to the
hydrogen bonding parameter $\tau$ which does not relate to the
tendency of monomers of the different kinds to avoid each other.
However, as it is shown by the considerations given in \cite{26},
\cite{27}, understanding of the whole picture of microphase
separation, including the temperature dependence of the structure
period, demands accounting the inhomogeneity in the distribution
of oligomers along homopolymer chains. Within the approach
developed, this is reached by means of the effective kinetic
energy (\ref{6}), with the mass fluctuating due to the temperature
dependence of hydrogen bonding. This dependence leads to the
reduction (\ref{36}) of the effective mass $m_{ef}$ that causes a
phase transition from stochastic to periodic distribution of the
oligomers along the homopolymer chain. However, if the critical
point is fixed usually by the condition $m_{ef}=0$ \cite{10}, in
our case the critical temperature $T_c$ relates to the finite
magnitude $m_{ef}={\bar m}/2$ of the effective mass which has a
singularity ${\rm d}m_{ef}/{\rm d}T=-\infty$ in the temperature
derivative (see Figure 2).

\begin{acknowledgements}
In this work, financial support by the Grant Agency of the Czech
Republic (grant GA\v{C}R 203/02/0653) is gratefully acknowledged.
\end{acknowledgements}

\appendix
\section{Derivation of a generic relation for microphase structure period}

Following \cite{2} we suppose the period of the microphase
structure to be determined by the minimum of the specific free
energy
\begin{equation}
f\equiv\frac{1}{V}\frac{F_{int}+F_{str}}{T};\quad V\equiv LS,\quad
L\equiv 2l+D \label{1a}
\end{equation}
related to the first term in Eq. (\ref{000}). Being the free
energy of the homopolymer--oligomer mixture, this term consists of
the interfacial and stretching components $F_{int}$, $F_{str}$
measured in the temperature units $T$ per the domain volume $V$
(according to Figure 1 $L$, $l$ and $D$ are the long space period,
the length of the oligomer tail and the thickness of the
homopolymer layer, respectively; $S$ is the domain surface area).

The interfacial free energy is stipulated by the loss of
conformational entropy caused by the localization of the
homopolymer chains within the interface of thickness $\Delta$. Due
to unfavorable interaction $\chi$ between the oligomer tails and
the polymer layer the chains form up loops containing segments of
number $\mathcal{N}_s\sim\chi^{-1}$ \cite{Helfand1971}. Then,
within the model of the random walk, the interface thickness is
estimated by the relations $\Delta^2\simeq\mathcal{N}_s b^2\sim
b^2/\chi$ where $b$ is the segment length. Respectively, the
interfacial free energy $F_{int}\simeq\mathcal{N}_l~T$ is
determined by the number $\mathcal{N}_l\simeq
S\Delta/\mathcal{N}_s b^3$ of the loops within the interface. As a
result, we obtain the estimation
\begin{equation}
F_{int}\sim\frac{\chi^{1/2}}{b^2}T~S. \label{1b}
\end{equation}

Another addition $F_{str}$ is caused by the stretching of the
surfactant side chains, whereas the stretching of the homopolymer
chains enlarges only the volume part of the free energy. This
addition is expressed by the simple equality
$F_{str}\simeq\mathcal{N}_c n_s F_1$ where the first factor
$\mathcal{N}_c\simeq DS/N b^3$ gives the number of chains per
layer, the second multiplier $n_s\simeq (b/\lambda)N$ is the
number of the oligomer molecules per chain of $N$ segments
($\lambda$ is period of the oligomers alternating along the chain)
and the last factor $F_1\sim (l^2/n b^2)T$ presents the free
energy of stretching a side chain of $n$ segments to a length $l$.
Combining the above multipliers, we find the estimation for the
total free energy of stretching
\begin{equation}
F_{str}\sim\frac{l^2 DS}{n b^4\lambda}~T. \label{1c}
\end{equation}

To derive the explicit expression for the dependence of the free
energy (\ref{1a}) on the layer thickness $D$ we need to use an
obvious condition $2lS\equiv\mathcal{N}_c n_s v_s$ where $v_s=n
b^3$ is the volume of surfactant molecule. As a result, we obtain
the relation
\begin{equation}
\frac{2l}{D}\equiv\frac{b}{\lambda}~n, \label{1d}
\end{equation}
according to which the period $\lambda$ defines the rest of
geometrical characteristics of the microphase separated structure.
Inserting Eqs. (\ref{1b}) --- (\ref{1d}) into Eq. (\ref{1a}), we
arrive to the final expression for the interfacial free energy:
\begin{equation}
f\sim\frac{\frac{\chi^{1/2}}{D}+\frac{n}{\lambda^3}D^2}{\left[1+(b/\lambda)n\right]b^2}
\label{1e}
\end{equation}
where numerical coefficients are dropped. The minimization
condition ${\partial f}/{\partial D}=0$ arrives at the
steady--state values of the homopolymer layer thickness and the
oligomer length
\begin{equation}
D\sim\frac{\chi^{1/6}}{n^{1/3}}\lambda, \quad
2l\sim\left(\chi^{1/6}n^{2/3}\right)b. \label{1f}
\end{equation}

It is convenient to express above results by means of the
dimensionless frequency of the oligomer alternating along the
homopolymer chain:
\begin{equation}
\omega\equiv\frac{2\pi}{\lambda/b}=\omega_{max}X;\quad
\omega_{max}\equiv 2\pi,~~X\equiv\frac{b}{\lambda} \label{1g}
\end{equation}
where $X$ is the averaged oligomeric fraction per homopolymer. So,
the long space period of the microphase separated structure takes
the form
\begin{equation}
L\equiv 2l+D\simeq
\chi^{1/6}n^{-1/3}\left(n+\frac{2\pi}{\omega}\right)b. \label{1h}
\end{equation}

\section{Calculations of convolution integrals}

\paragraph{Self--energy functions}
Calculations of the self--energy functions (\ref{41}), (\ref{42})
lead to a rather tedious procedure due to the convolution
integrals. To demonstrate the line of these calculations we
consider in details the simplest integral related to the first
term in Eq. (\ref{42}):
\begin{eqnarray}
\Sigma_{-}^{\mu}(\nu)=\mu^2\int\frac{{\rm d}\nu_1}{2\pi} S(\nu_1)
G_{-}(\nu-\nu_1). \label{a2}
\end{eqnarray}
Making use of the expressions (\ref{29}) for the structure factor
$S$ and Green function $G_{-}$, where the frequency dispersed
parameter $\tau(\nu)$ is replaced by its bare magnitude $\tau$,
arrives at the convolution integral
\begin{eqnarray}
\Sigma^{\mu}_{-}(\nu)=\frac{\mu^2}{2\pi n_c^3}\int \frac{{\rm
d}\nu_1}{(\omega_s^2+\nu_1^2)\left[\omega_s-{\rm i
}(\nu-\nu_1)\right]}\label{a4}
\end{eqnarray}
where a characteristic frequency $\omega_s\equiv \tau/n_c$ is
introduced. This integral has the poles $\pm{\rm i}\omega_s$ and
$\nu+{\rm i}\omega_s$ (see Figure 7a).
\begin{figure}[!h] \label{polus}
\epsfig{file=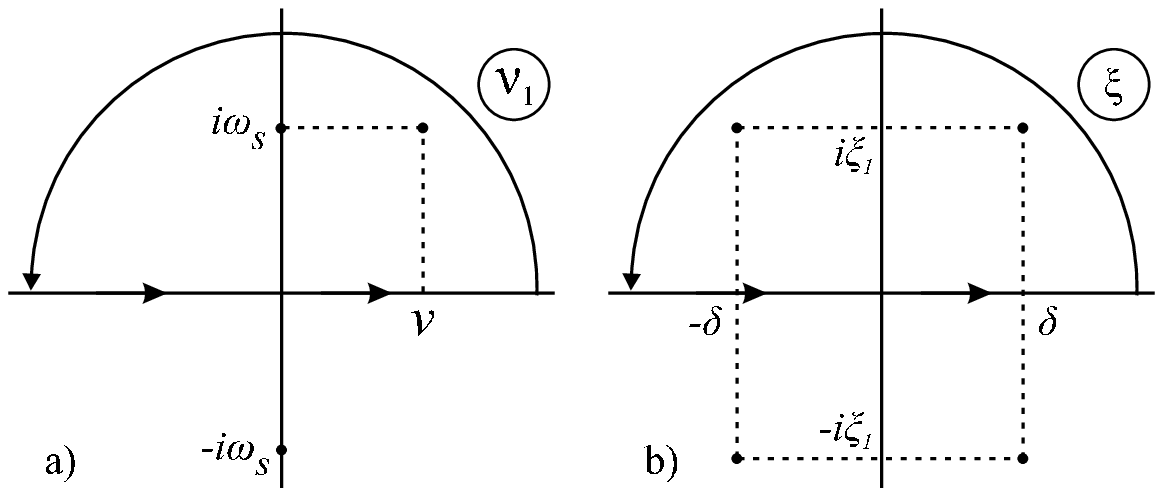,width=80mm,bbllx=17mm,bblly=204mm,bburx=142mm,bbury=265mm}
\caption{Poles of the convolution integral in Eqs. (\ref{a4}) (a)
and (\ref{xi}) (b). $\delta\equiv\sqrt{\xi_2^2-\xi_1^2}$.}
\end{figure}
In accordance with the theory of residues \cite{book poles}, the
integral in Eq. (\ref{a4}) is reduced to sum over two of these
residues that locate in upper half--plane of the complex frequency
$\nu_1$:
\begin{equation}\label{poles}
2\pi{\rm i}\left[\frac{-\rm{i}}{2{\rm
i}\omega_s\cdot(-\nu)}+\frac{-\rm{i}}{\nu\cdot(\nu+2{\rm
i}\omega_s)}\right]
\end{equation}
where terms in the square brackets relate to the poles ${\rm
i}\omega_s$ and $\nu+{\rm i}\omega_s$, respectively. After a
simple algebra this expression yields
\begin{eqnarray}
\Sigma_{-}^{\mu}=\frac{\mu^2}{\tau^2
n_c}~\frac{1+\frac{\rm{i}}{2}\xi}{4+\xi^2},\quad
\xi\equiv\frac{\nu}{\omega_s}. \label{a10}
\end{eqnarray}
Analogously, the rest of convolution integrals is calculated
giving final form of Eqs.(\ref{41}), (\ref{42}):
\begin{eqnarray}
\Sigma=\frac{\mu^2}{2\tau^3 n_c}~\frac{1}{4+\xi ^2}+
\frac{\lambda^2}{8\tau^4 n_c^2}~\frac{1}{9+\xi ^2},
\label{a11} \\
\Sigma_{\pm}=\frac{\mu^2}{\tau^2 n_c}~\frac{1\mp{{\rm i}\over
2}\xi }{4+\xi ^2}+ \frac{\lambda^2}{8\tau^3 n_c^2}~\frac{3\mp{\rm
i}\xi }{9+\xi ^2}. \label{a11a}
\end{eqnarray}

\paragraph{Renormalization mass parameter}
Explicit form of the renormalization mass parameter (\ref{35}) is
determined by the structure factor (\ref{39a}) and Green function
(\ref{45}) with the effective mass (\ref{36}) and parameter
$\tau(\nu)$ being replaced by bare $\tau$:
\begin{equation}
\Delta=\frac{\sigma^2}{\pi n_c
m_{ef}}\int\frac{(\xi_1-\xi_0^{-1}\xi^2)\xi^2{\rm
d}\xi}{(\xi^2-2{\rm i}\xi_1\xi-\xi_2^2)(\xi^2+2{\rm
i}\xi_1\xi-\xi_2^2)} \label{xi}
\end{equation}
where one denotes
\begin{equation}
\begin{split}
\xi_0\equiv &\frac{16
m_{ef}\tau}{n_c}\left(\frac{\mu^2}{4}+\frac{\lambda^2}{3^4\tau
n_c}\right)^{-1},\\
\xi_1\equiv &\frac{n_c^2}{2 m_{ef}\tau}\left[1+\frac{1}{8\tau^3 n_c}\left(\mu^2+\frac{\lambda^2}{9\tau n_c}\right)\right],\\
\xi_2^2\equiv &\frac{n_c^2}{m_{ef}\tau}\left[1-\frac{1}{4\tau^3
n_c}\left(\mu^2+\frac{\lambda^2}{6\tau n_c }\right)\right].
\label{def}
\end{split}
\end{equation}
The integral in Eq. (\ref{xi}) has the pole structure that is
shown in Figure 7b. As above, the sum over residues located in the
upper half--plane of the complex frequency $\xi$ yields the
integral value
\begin{equation}
\frac{\pi}{2}\left(1-\frac{\xi_2^2-4\xi_1^2}{\xi_0\xi_1}\right).\label{a12}
\end{equation}
With accounting the notices (\ref{def}) and keeping only the terms
of the second order of smallness over parameters $\mu$ and
$\lambda$, one obtains the final expression (\ref{46a}).

\end{document}